\documentclass[11pt]{article}
\usepackage{bm,amsmath,color,dsfont, amssymb}
\usepackage[titletoc,toc,title]{appendix}
\usepackage{graphicx}
\usepackage[section]{placeins}
\usepackage{esint}
\usepackage{adjustbox}
\usepackage[hidelinks]{hyperref}
\usepackage{caption}
\usepackage{cleveref}
\usepackage{subcaption}
\usepackage{tabularx}
\usepackage{tgtermes} 
\usepackage{relsize}
\usepackage{booktabs}
\usepackage{wrapfig}
\usepackage{siunitx}
\usepackage{bm}
\usepackage{xcolor}
\usepackage{booktabs}
\usepackage{multirow}
\usepackage{float}

\usepackage{authblk}
\begin{document}

\title{Navigating through Economic Complexity:\\
Phase Diagrams \& Parameter Sloppiness}
\author{Jean-Philippe Bouchaud,\\ Capital Fund Management \& Académie des Sciences}
\date{November 2024}
\maketitle

\begin{abstract}
We argue that establishing the phase diagram of Agent Based Models (ABM) is a crucial first step, together with a qualitative understanding of how collective phenomena come about, before any calibration or more quantitative predictions are attempted. Computer-aided {\it gedanken} experiments are by themselves of genuine value: if we are not able to make sense of emergent phenomena in a world in which we set all the rules, how can we expect to be successful in the real world? ABMs indeed often reveal the existence of Black Swans/Dark Corners i.e. discontinuity lines beyond which runaway instabilities appear, whereas most classical economic/finance models are blind to such scenarii. Testing for the overall robustness of the phase diagram against changes in heuristic rules is a way to ascertain the plausibility of such scenarii. Furthermore, exploring the phase diagrams of ABM in high dimensions should benefit enormously from the identification of ``stiff'' and ``sloppy'' directions in parameter space. 

This paper stems from my contribution to G. Dosi's 70th Festschrift in Pisa, November 2023.
\end{abstract}

\section{Introduction}

\subsection{From micro-rules to macro-behaviour}

Inferring the behaviour of large assemblies from the behaviour of its elementary constituents is arguably one of 
the most important problems in a variety of different disciplines: physics, 
material sciences, biology, computer sciences, sociology and,
of course, economics and finance. It is also a notoriously hard problem. Statistical physics has developed in the last 150 years precisely to grapple with the complex relation between microscopic entities and macroscopic behaviour, with the understanding of phase transitions and collective phenomena as trophy achievements \cite{Sethna}. 

Clearly, when interactions are absent or small enough,
the system as a whole merely reflects the properties of individual
entities. This is the canvas of traditional macro-economic approaches. Economic
systems are assumed to behave as if populated by a collection of identical, non-interacting agents that represent the average properties of heterogeneous agents -- the so-called ``Representative Agent''. This average agent is furthermore endowed with formidable cognitive and computational abilities, as if the irrationality and behavioural biases of individual agents would average out. 

However, we know (in particular from physics) that discreteness, heterogeneities and/or interactions can lead to totally unexpected phenomena. Think for example of super-conductivity or super-fluidity\footnote{See e.g. Ref.~\cite{Balibar} for the fascinating history of the discovery of super-fluidity.}: 
before their experimental discovery, it was simply beyond human
imagination that individual electrons or atoms could ``conspire'' to create a collective state that can flow without friction. Micro and macro behaviour do not coincide in general: in fact, genuinely {\it surprising} behaviour can emerge through aggregation \cite{anderson_more, bouchaud_cdf}. 

From the point of view of economic theory, this is
interesting because financial and economic history is strewn with bubbles, crashes, crises and upheavals of all sorts. These are very hard to fathom within a Representative Agent framework~\cite{Kirman}, within which crises can only appear due to large aggregate shocks, when in fact small local shocks can trigger large systemic effects when heterogeneities, interactions and network effects are taken into account~\cite{Brock,Complexity,JPB,Kirman_complex,Arthur,PNAS,Dosi_complex, Dosi_book, Dessertaine}. 

Because these effects are difficult to account for within analytical models, numerical simulations of ``Agent-Based models'' (ABMs) are needed, and have a long history in physics, epidemiology, biology, and more recently macroeconomics, see e.g. \cite{ABM-collective, Epstein, Foley} and references below. 

These models are extremely versatile because 
any possible behavioural rules, interactions,
heterogeneities can be taken into account. In fact, these models are so versatile that they suffer from the ``wilderness of
high dimensional spaces'' (paraphrasing Sims \cite{Sims}). The number of 
parameters and explicit or implicit choices of behavioural rules is so large ($\sim 10$ in the simplest ``Mark-0'' model, see below, but often in the hundreds) that  the results of the model may appear unreliable and arbitrary, and the calibration of the parameters is an hopeless (or highly unstable) task. 

Mainstream ``Dynamic Stochastic General Equilibrium'' models (DSGE), on the other hand, are simple enough to lead to closed form analytical results, with simple narratives and well-trodden calibration avenues~\cite{gali}, although full-blown DSGE models also contain several dozens of parameters, see e.g. \cite{Rohe}.
{
In spite of their unrealistic character, these models appear to perform 
satisfactorily in `normal' times, when fluctuations are small. However, they become deeply flawed in times of economic instability \cite{Buiter}, suggesting different assumptions are needed to understand what is observed in reality.} But even after the 2008 crisis, traditional equilibrium models, augmented with frictions and heterogeneities of different kinds, are still favoured by most economists, both in academia and in institutional and professional circles, see e.g. \cite{blanchard, HANK}. For example, the recent period of high inflation (2021 -- 2023) has been analyzed through the lens of DSGE/New Keynesian models by prominent economists \cite{Reis}, stirring a heated debate \cite{Rudd,Stiglitz, blanchard_bernanke}.\footnote{For an extension of DSGE models to account for interaction between agents, see \cite{PNAS}.} ABMs are only slowly starting to be considered as a viable alternative \cite{haldane,Poledna, Farmer_book, Farmer_new} -- see \cite{Eurace, Eurace2,SantAnna, Dosi,Lagom,Caiani} for early macroeconomic ABMs,  \cite{Roventini} for an enlightening discussion on the debate between traditional DSGE models and  ABMs, and \cite{Fagiolo1,Fagiolo2,Kirman2,Caballero} for further insights.

\subsection{A methodological manifesto}

At this stage, it seems to us that some clarifications are indeed needed,
concerning both the objectives and methodology of Agent-Based Models. 
ABMs do indeed suffer from the wilderness of high dimensional spaces. One possibility, advocated in particular by D. Farmer \& collaborators \cite{Farmer_covid, Farmer_book, Farmer_new}, is to use as much micro-data and domain knowledge as possible, in particular to initialize as faithfully as possible the state of the economy, and then to run the model for a short enough amount of time so that the future state of the economy can be inferred from relatively constrained input-output rules for production, consumption and depreciation. Such an ``educated extrapolation'' approach has led to remarkable results concerning the fate of the UK economy a few months after the COVID lockdown. Counter-factuals can be run, providing an immensely useful tool for policy makers, at least on the short run \cite{Poledna, Farmer_covid}. 

The long time behaviour of such models is however not (yet) as convincing, and the sensitivity to parameters is expected to grow as the time horizon expands. Are these synthetic economies intrinsically stable or can endogenous crises appear? Can inflation bouts, such as the one we lived through in the years 2022 -- 2023, be dynamically generated as a result of massive fiscal policy during COVID? What are the possible long term effects of fiscal/monetary policies? More generally, what are the possible emergent behaviours that can be expected given the set of micro-rules defining an ABM?

In this respect, statistical physics offers a key concept: the {\it phase 
diagram} in parameter space~\cite{Sethna}. A classic example is the 
phase diagram of usual substances as a function of two parameters,
temperature and pressure. The generic picture is that:
\begin{itemize}
    \item The number of distinct phases is usually small, in any case much smaller than the number of parameters (e.g. three in the previous example: solid, liquid, gas).
    \item Well within each phase, the properties are qualitatively similar  and small changes of parameters have a small effect. Macroscopic (aggregate) properties do not fluctuate any more for very large systems and are robust against changes of microscopic details -- for example the shape and chemical nature of molecules composing the substance.  
\end{itemize} 
This is the ``nice'' scenario, where the dynamics of the system can be described in terms of a small number of relevant variables (not necessarily obvious from the outset) with some effective parameters that parsimoniously encode the microscopic details. But other scenarios are of course possible. For example, if one sits close to the boundary between two phases, fluctuations can remain large even for large systems. Small changes of parameters can then 
radically change the macroscopic behaviour of the system, corresponding to ``parameter chaos''. Interestingly, there may be plausible mechanisms driving the system close to criticality (like Self Organized Criticality~\cite{Bak, Bouchaud_SOC}). Alternatively, in some cases like spin-glasses~\cite{SG}, whole phases are critical and are characterized by so-called ``parameter chaos''. 

In any case, before any calibration attempt, the very first step in exploring the properties of an Agent-Based model should be to identify the different possible phases in parameter space and the location of the phase boundaries \cite{gualdi}. In order to do this, numerical
simulations turn out to be very helpful \cite{Buchanan,Foley,Farmer_book}. Indeed, within each phases aggregate behaviour usually quickly sets in, even for small sizes. As we explain further below, only a handful of parameters (or combination of parameters) turn out to be crucial, while others are found to play little role. This is useful to establish a qualitative {\it phenomenology} of the model -- what kind of behaviour can the model reproduce,
which basic mechanisms are important, which effects are potentially missing?
This first, qualitative step allows one to unveil the ``skeleton'' of the ABM.
Simplified models that retain most of the phenomenology can then be constructed
and perhaps solved analytically, enhancing the understanding of the important
mechanisms, and providing some narrative to make sense of the observed effects \cite{gualdi,gualdi_prl}. In order to avoid getting lost in the wilderness of high dimensional models, calibration of an ABM using real data should only start to after such a qualitative investigation has been performed, using specific tools that we will describe below. The phase diagram of
the model then allows one to restrict the parameters to regimes that lead to ``reasonable'' outcomes.

\subsection{Telescopes for the mind}

While it is obviously hard to perform controlled experiments in a macroeconomic context, one can argue that ABMs provide computer-aided {\it gedanken} experiments of genuine value, that help theorists train their intuition by forcing them to think about the often unexpected results of numerical simulations -- much like Galileo's telescope helped deciphering the motion of the planets. This is what Mark Buchanan called ``telescopes for the mind'' \cite{Buchanan}: If we are not able to make sense of an emergent phenomenon in a world in which we set all the rules, how can we expect to be successful in the real world? ABMs can reveal the existence of ``Black Swans'' or ``Dark Corners'' \cite{Blanchard}, i.e. discontinuity lines beyond which runaway instabilities appear, whereas most classical economic/finance models are blind to such scenarii. 

Hence, even stylized ABMs that are not meant to be accurately calibrated on real data can provide useful insights by generating qualitatively plausible scenarios and counter-factuals. In fact, one should abandon the ``pretense of knowledge''~\cite{Caballero} and false sense of control provided by simple linear models and opt for a more qualitative, scenario-based approach to macroeconomic phenomena, with emphasis on non-linear mechanisms, feedback loops, etc. rather than on precise-looking, but possibly misleading numbers. Hence, even if stylized macro-ABMs (like the Mark-0 ABM discussed in the next section) turn out to be little more 
than a methodological exercise, we strongly believe that it should be part of the curriculum of economics students and the toolkit of policy makers, if only as an inspiring scenario generator, or “telescope for the mind” \cite{Buchanan}, especially in times of great modeling uncertainty during which it is crucial to be at least
“roughly right” and avoid being blindsided by in-name-only Black Swans. 

\section{Phase Diagrams, Tipping Points \& Surprises in Macro-ABMs}

In this section, we want to illustrate the general philosophy outlined in the previous section using the ``Mark-0'' model. The Mark-0 ABM is a simplified model of a closed macroeconomy that nonetheless generates a wide variety of plausible scenarios, from stable low unemployment and inflation, to endogenous crises that may oscillate regularly or punctuate long periods of recovery, or even to runaway inflation. The concept of phase diagram has been extremely useful to map out the different emergent phenomena that can emerge from the micro-rules of Mark-0.  

The Mark-0 economy is made of firms and households. While the latter sector
is represented at an aggregate level (therefore neglecting the possibly important role of 
wage and wealth inequalities), firms are heterogeneous and treated individually.
Each firm $i$ produces $Y_i(t)$ perishable goods that
it attempts to sell at price $p_i (t)$. It needs a number of $N_i (t) = Y_i (t)/\zeta_i$ of employees to produce (where $\zeta$ is a measure of productivity), and pays a wage $W_i (t)$. 

The demand $D_i$ for good i depends on the global consumption budget of households, itself determined as a fraction of the household savings. $D_i$ is a decreasing function
of the asked price $p_i (t)$, with a price sensitivity parameter that can be tuned.

To update their production, price and wage policy, firms use reasonable ``rules of
thumb” introduced by Delli Gatti et al. \cite{MarkIref,MarkIbook}. 
For example, production is decreased and employees are made
redundant whenever $Y_i > D_i$, and vice-versa. The adjustment speed can however be
asymmetric, i.e. the ratio $R$ of hiring adjustment speed to firing adjustment speed
is not necessarily equal to one, for example because of labour laws. This turns out to
be one of the most important control parameter that determines the fate of the overall
economy.

When the Mark-0 economy is set in motion, it soon becomes clear that some firms
have to take up loans in order to stay in business. One therefore immediately has to
add further rules for this to take place. We let firms freely accumulate a total debt up to a threshold that is a multiple $\Theta$ of total expected sales. Beyond this threshold, the firm is declared bankrupt (its debt is then repaid partly by households and partly by surviving firms, such that there is no creation of money out of thin air). 

From this point of view the parameter $\Theta$ determines the maximum credit
supply available to firms. Fixing the value of $\Theta$ plays the role of a primitive monetary
policy, since the total amount of money circulating in the economy (‘broad money’)
directly depends on $\Theta$ \cite{gualdi}. When $\Theta$ = 0, no debt is allowed (zero
leverage), while when $\Theta \to \infty$, firms have no limits on the loans they need to continue
business.

\begin{figure}
\centering
\includegraphics[scale=0.3]{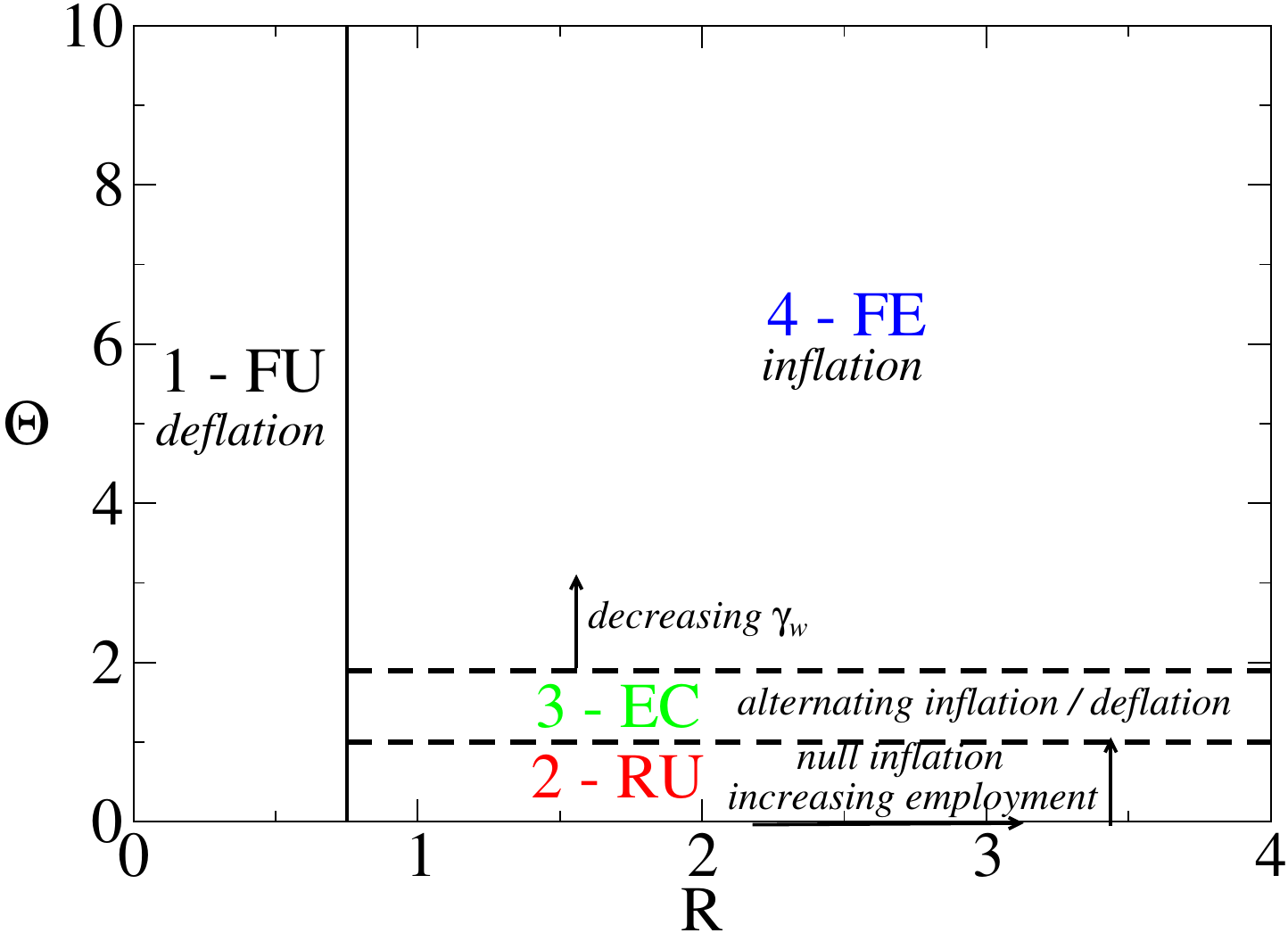}
\includegraphics[scale=0.3]{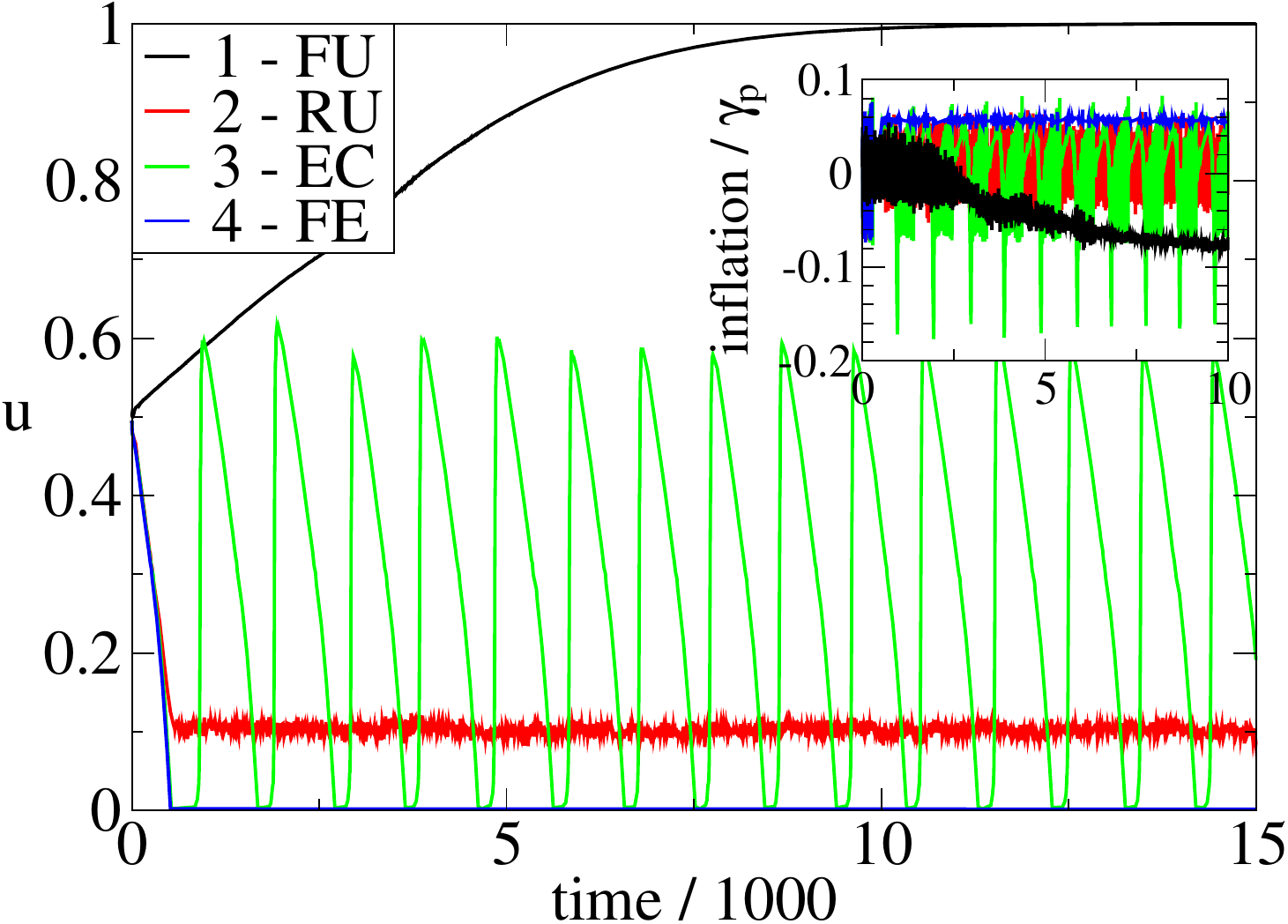}
\caption{\small{
({\it Left})
Phase diagram in the $R-\Theta$ plane of the basic Mark-0 model as obtained
in~\cite{gualdi} with wage update. There are four distinct phases separated by critical lines. The Full Employment (FE) phase ($R > R_c$, $\Theta$ large) is characterised by positive average inflation, while there is deflation in the Full Unemployment (FU) phase ($R < R_c$). Endogenous Crises (EC) are characterised by alternating cycles of inflation and deflation, and occur for $R > R_c$, $\Theta$ intermediate. Finally, $R > R_c$, small $\Theta$ correspond to
a region of small inflation and Residual Unemployment (RU).
The location of phase boundaries is only weakly affected by the choice of the other parameters of Mark-0, see~\cite{gualdi}.
({\it Right})
Typical trajectories of the unemployment rate $u(t)$ for each of the phases. In the inset, the price variations are shown, displaying either inflation (in the FE phase) or deflation (in the FU phase). $\gamma_p^{-1}$ (resp. $\gamma_w^{-1}$) sets the characteristic time scale for price adjustments (resp. wages) in the model, see \cite{karl_inflation}. The surprising occurrence of endogenous oscillations in the EC phase can be fully understood analytically, see~\cite{gualdi_prl}.
}}
\label{fig:PD_wages}
\end{figure}

While there are several other parameters needed to define completely Mark-0 (see e.g. \cite{karl_inflation} for the most recent version), the numerical investigation of \cite{gualdi} 
has suggested that parameters $R$ and $\Theta$ play a particularly important role in determining the 
aggregate behaviour, as illustrated in phase-diagram of the shown in Fig. 1. 
As expected from the general intuition gained from the study of physical systems, such phase diagram is found to be extremely robust against both details of the model specification and the value of the other parameters, which leave the qualitative emergent behavior unchanged but affect the precise location of the phase boundaries. In the bare-bone version of Mark-0, four distinct phases can be identified \cite{gualdi,karl_sloppy}:
\begin{itemize}
    \item When $\Theta = \infty$ the economy is characterized by two distinct phases separated by a
discontinuous phase transition as a function of the parameter $R$. When $R < R_c$ (fast downward production adjustments), one finds at long times a collapse of the economy towards a deflationary/low demand/full unemployment state (FU). For $R > R_c$ (fast upward production adjustments), on the other hand, the long run state of the economy is characterized by a positive inflation/high demand/full employment phase (FE).
\item  When $\Theta < \infty$ the above description carries through but must be refined to allow for the appearance of three sub-phases for $R > R_c$:
\begin{enumerate}
    \item a full employment and inflationary phase for high values of $\Theta$ (the FE phase,
similar to the $\Theta = \infty$ case);
\item a phase for intermediate values of $\Theta$ characterized by high employment and
inflation on average, which is surprisingly disrupted by ``endogenous
crises” (EC), that temporarily bring deflation and high unemployment
spikes;
\item a phase with zero inflation and residual unemployment for small $\Theta$ (the RU
phase), where the impossibility to obtain loans creates a positive stationary
level of bankruptcies.
\end{enumerate}
\end{itemize}
Note that the existence of these endogenous crises is a genuine ``surprise'', i.e. an effect that appears at the aggregate level, that was hard to anticipate from the simple rules at the micro-level. Such endogenous oscillations can in fact be understood fully analytically \cite{gualdi_prl} and result from very mild assumptions about the destabilizing feedback mechanisms present in the Mark-0 economy. Interestingly, such macro-economic oscillations are mathematically akin to spontaneous oscillations taking place in the brain or in the flashing of fireflies -- which are also surprising emergent phenomena that took decades to account for \cite{strogatz}.

Although far from perfect, Mark-0 contains plausible ingredients that are most probably
present in reality. For example, our model encodes in a schematic manner the consumption
behavior of households facing inflation, that is in fact similar to the standard Euler equation
for consumption in general equilibrium models \cite{gali}. It is however important to stress that
the behavioral rules that define Mark-0 are only reasonable when the economy
behaves normally, and we believe that they correctly describe how such a normal state may
become unstable. However, once the instability happens and the economy truly collapses, it
is of course unreasonable to expect that agents will keep acting according to the same rules,
in particular the Central Bank and other institutions. Including monetary and fiscal policies within the Mark-0 framework is of course possible and was discussed in \cite{monetary, inflation, sharma, karl_inflation}, and we briefly discuss the results of such investigations in the context of phase diagrams in the next section.  

\section{Monetary policy \& Inflation anticipation: More Tipping Points}

In the original specification of Mark-0, no feedback between realized inflation and decisions of households and firms was considered, nor any Central Bank (CB)  monetary policy -- such effects were considered in later versions of the model, and can again be analyzed within the language of phase-diagrams. Let us give three representative examples: (a) unintended consequences of monetary policy; (b) transition between V-shape and L-shape recovery after a COVID-like shock; (b) feedback induced-hyper inflation.

We first describe cursorily how inflation and interest rate affect the behaviour of the Mark-0 economic agents. We posit that agents anticipate future long term inflation as an average between past {\it realized} inflation and the inflation target of the Central Bank. When trust in the CB is high, inflation expectation are anchored and high weight $\tau^T$ is given to the inflation target. Trust itself can be made dynamical, i.e. $\tau^T \to \tau^T(t)$ and may decrease when realized inflation is persistently off-target (``floating trust''). 

The CB sets the interest rate $\rho$ using the standard Taylor rule, i.e. it increases $\rho$ from the baseline rate $\rho^\star$ when inflation or employment is above target, with certain Taylor parameters $\alpha_{\pi,e}$. The increase of $\rho$ has a direct effect on the consumption of households, as savings are favoured (this is the analogue in Mark-0 of the so-called Euler equation in DSGE models). Conversely, increased anticipated inflation leads to an increase of consumption. Similarly, the balance between interest rate and anticipated inflation directly impacts the policy of firms in terms of hiring/firing. For example, strongly indebted firms will fire more rapidly when interest rates increase. Anticipated inflation also directly impact the price and wage setting policies of firms through, respectively, a market power coefficient $g_p$ and a bargaining power coefficient $g_w$ -- see \cite{karl_inflation} for a full discussion. 

When these ingredients are added to Mark-0, one can perform different {\it in silico} experiments and classify the results in a synthetic way in phase diagrams. 

\begin{figure}
\centering
\includegraphics[scale=0.8]{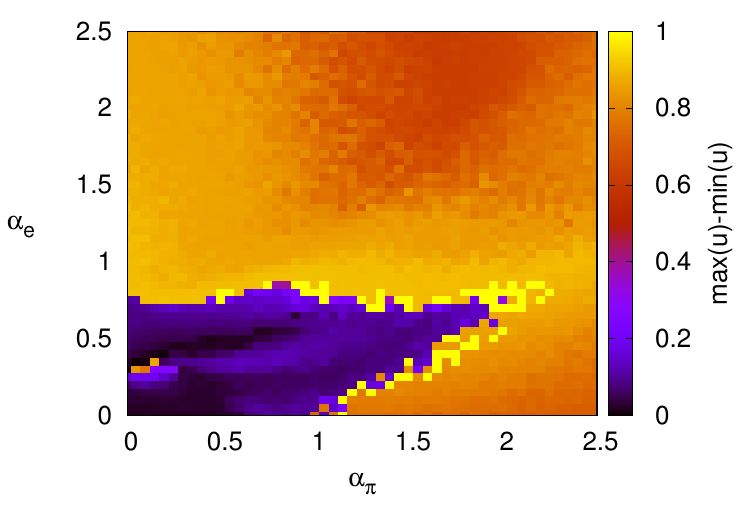}
\caption{\small{
Policy performance in the Taylor parameters plane $(\alpha_\pi,\alpha_\varepsilon)$. Color map 
of the amplitude of the business cycle, measured as $\max_t(u)-\min_t(u)$. Yellow/orange regions correspond to unstable economies with crises of large amplitude. 
As one can see the policy is effective as long as it is not too aggressive, with a sharp transition to a regime
where it become detrimental. The closer the ``natural'' economy (without CB) to a phase boundary, the more destabilising the policy. From Ref. \cite{monetary}.
}}
\label{fig:ap-ae}
\end{figure}

\paragraph{a. Unintended consequences of monetary policy.} We show in Fig. 2 that hyper-reactive monetary policies (with large Taylor parameters $\alpha_{\pi,e}$) can in fact destabilize the economy and induce periods of high unemployment. This result is completely at odds with the outcome of rational-expectation, DSGE models for which {\it the monetary authority should respond to deviations of inflation and the output gap from their target levels by adjusting the
nominal rate with “sufficient strength”; [...] it is the presence of a “threat” of a strong response by the monetary authority to an eventual deviation of the output gap and inflation from target that suffices to rule out any such deviation in equilibrium (!)} \cite{gali}. Such a conclusion appears to be strongly blighted when agents follow more naive rule of thumb strategies. Placing faith in the rationality of agents can be quite dangerous. In fact, as expressed beautifully by Giuliano da Empoli in his book ``Le Mage du Kremlin'': {\it Il n'y a rien de plus sage que de miser sur la folie des hommes}.

\paragraph{b. V-shape vs. L-shape recovery after COVID, and policy induced inflation} The impact of the COVID pandemic on the economy has been through both a fall in consumption and a simultaneous loss in productivity. To implement such Covid–induced dual consumption and productivity
shocks, we proposed to reduce the productivity factor $\zeta$ by an adjustable quantity $\Delta \zeta$ and the consumption propensity of households $c$ by $\Delta c$ \cite{sharma}. As shown in Fig. 3, for small enough  $\Delta c$ and/or $\Delta \zeta$, one observes a fast ``V-shape'' recovery when lockdown is over, as expected when the shock is mild enough not to dent the financial health of the firms. Stronger shocks can however lead to a permanent dysfunctional state (L-shape), with high unemployment, falling wages and savings, and a high level of financial fragility and bankruptcies. Interestingly, there is an abrupt, first order transition line (tipping point) beyond which crises have a very large probability to become permanent, with high levels of unemployment.

\begin{figure}[t]
    \centering
    \includegraphics[width=\linewidth]{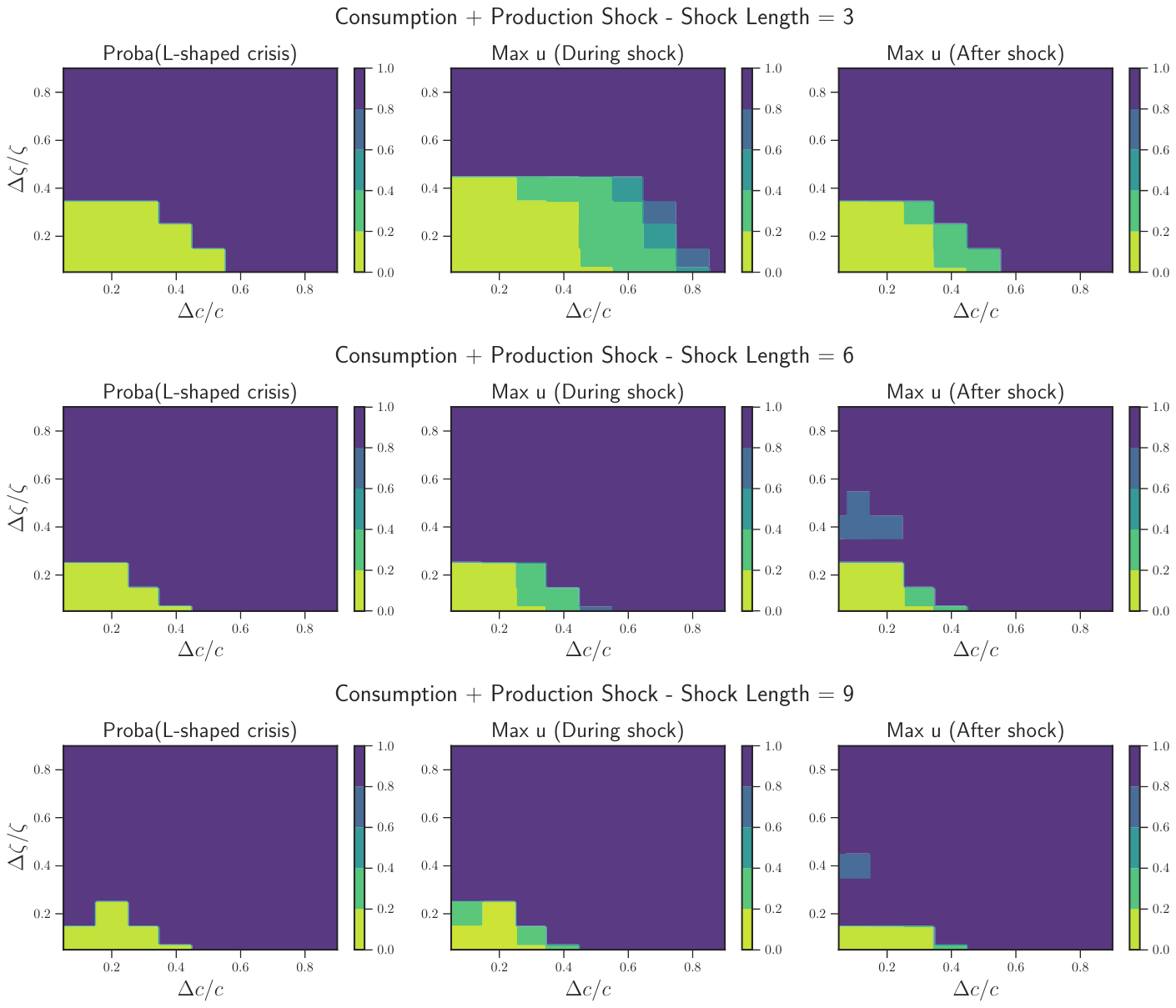}
    \caption[Phase diagrams in the $\Delta c / c$-$\Delta \zeta / \zeta$ plane for different shock lengths]{\small{Phase diagrams in the  $\Delta c / c$ - $\Delta \zeta / \zeta$ plane for different shock lengths. 
      \textit{Top Row:} Shock lasting for 3 months. The region of parameter space with no L-shaped crisis is quite large allowing for strong consumption shocks ($\Delta c/c \lesssim 0.5$) and mild productivity shocks ($\Delta \zeta/\zeta \lesssim 0.3$). Note that for such a short shock, the effects on unemployment are seen after the shock has ended. \textit{Middle Row:} Shock lasts for 6 months. A decrease in the region of no-crisis is observed. Mild shocks ($\Delta c/c \sim 0.4$) can also lead to prolonged crises. During the shock itself, extremely high rates of unemployment can be seen. \textit{Bottom Row:} Shock lasts for 9 months. Only for extremely mild shocks does the economy not undergo a prolonged crisis. From Ref. \cite{sharma}. }}
    \label{fig:phase_diagrams}
  \end{figure}

In the real economy, the precise location of such tipping points is hard to estimate. We are again led to conclude that governments and other institutions should err on the side of caution and do ``whatever it takes” to prevent the economy from falling into a bad state, supporting firms and households, and stimulate a rapid recovery. 

However, when implementing such policies within Mark-0, for example in the form of easy access of firms to credit, we found that when policy is successful, post-crisis inflation was significantly increased compared to the pre-crisis period, a predicament that did indeed occur some months after our paper was published \cite{sharma}. It would of course be preposterous to claim that we had predicted the mid-2021 to mid-2023 inflation spike that took Central Bankers by surprise  and befuddled DSGE pundits \cite{Reis,blanchard_bernanke}. Still, we believe that at least part of the story was correctly captured by Mark-0 while being mostly out-of-reach from equilibrium models. This again illustrates the main upside of stylized ABMs: their ambition is not to provide precise predictions based on a fully calibrated model, but rather a tool for decision makers to help them apprehend
different possible outcomes and anticipate unintended consequences and potential
counter-intuitive impacts of their policies.    

\paragraph{c. The risk of a hyperinflation spiral.}

We can interpret the crucial role of wage bargaining power $g_w$ and market pricing power $g_p$ in light of the dangers of a hyperinflation episode resulting from a wage-price spiral. Alvarez et al. \cite{alvarez} in 2022 voiced concern that hyperinflation may occur if firms increase wages in response to higher inflation,
leading to an increase in purchasing power and ultimately feeding into a wage-price
spiral in the current macroeconomic environment. Such feedback loop is influenced
by the indexation of prices and wages to firms’ inflation expectations \cite{holland}, i.e. by the value of parameters $g_p$ and $g_w$ in our model. In the simplest case where wage bargaining power and market power are equal ($g_p=g_w$), the economy reaches a stable inflationary state, marked by cyclical fluctuations due to mismatches in supply and demand. When indexation
is weak ($g_p=g_w < 1$), neither wages nor prices fully incorporate inflation expectations, and fluctuations in the inflation rate due to mismatches in demand and supply are dampened. 

Conversely, for strong indexation ($g_p=g_w > 1$), the economy may enter a state of hyperinflation, though as wages and prices increase equally fast,
there are no real effects (omitting ``menu
costs”). Only when $g_p \neq g_w$ does the economy collapse.

Introducing monetary policy implies that hyperinflation can be staved off due to
the anchoring of inflation expectations. In the case of anchored expectations, inflation remains stable until a critical point $g_p=g_w < g^\star$ where $g^\star > 1$ depends on the commitment of the Central Bank as well as the strength of expectation anchoring to the its target (see Fig. 4). However, in the case $1 < g_p=g_w < g^\star$ inflation is only superficially stable. Indeed, a strong enough shock triggers a loss of trust in the Central Bank, tipping the economy into hyperinflation. This is precisely the self-reflexive mechanism dreaded by Central Bankers, that may lead to hyperinflation when expected future inflation
is dominated by past realised inflation, rather than by the CB inflation target.

\begin{figure}[t]
    \centering
    \includegraphics[width=\textwidth]{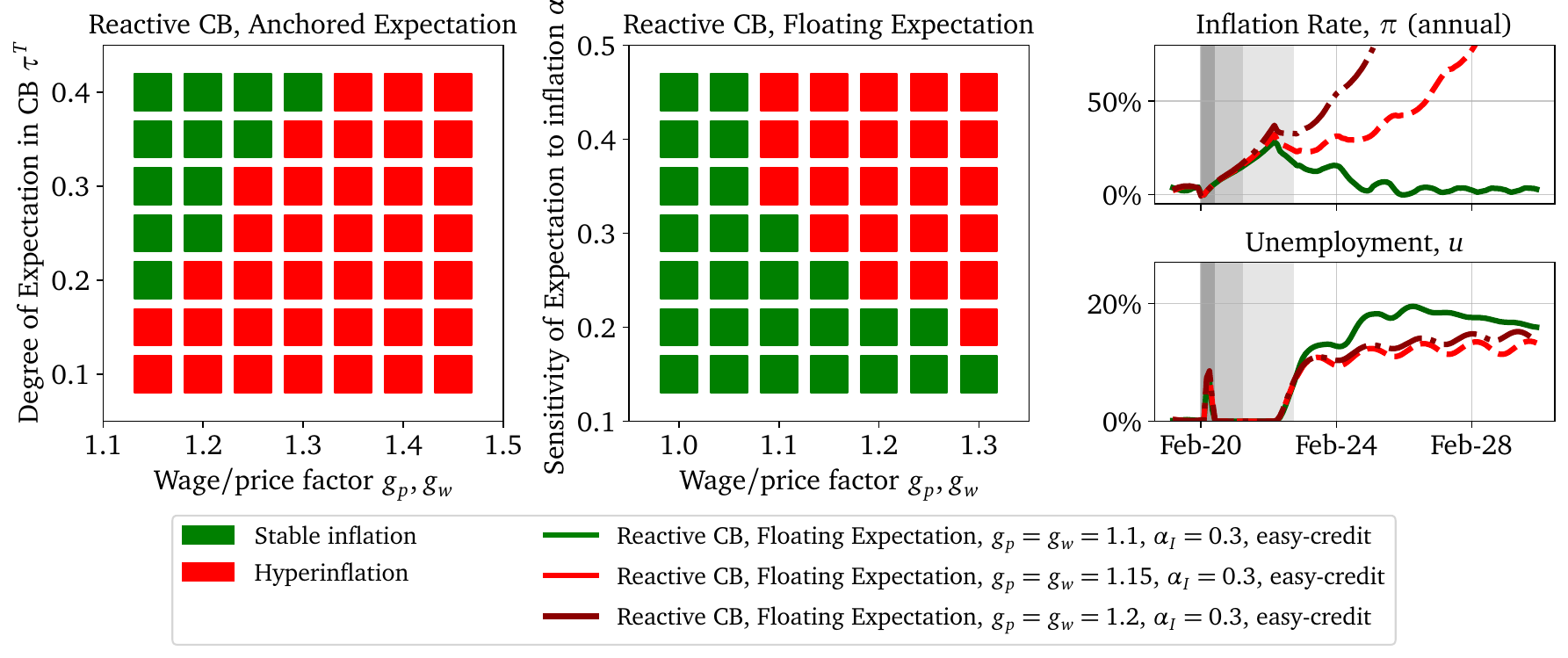}
    \caption{\small{{Hyperinflation tipping points.} {\it Left:} Stable inflation vs. Hyperinflation in the plane $g_p=g_w$, $\tau^T$ (where $\tau^T$ measures the weight given to the CB inflation target) in the Anchored Expectation scenario and in absence of exogenous shocks. Note that inflation can be stable even when $g_p=g_w > 1$ for strong enough anchoring. {\it Middle:} Stable inflation vs. Hyperinflation in the plane $g_p=g_w$, $\alpha_I$ (where $\alpha_I$ measures the speed at which agents lose trust in the CB when inflation is off-target, see Ref. \cite{karl_inflation}) in absence of exogenous shocks. {\it Right:} inflation rate (top) and unemployment (bottom) as a function of time, in the Floating Expectation case, $\alpha_I=0.3$ and different values of $g_p=g_w$, after the COVID shocks. When indexation is too strong, an hyperinflation regime sets in at the end of the energy price shock. From \cite{karl_inflation}.}}
    \label{hyperinflation_phase}
\end{figure}

\section{Parameter sloppiness: navigating through large dimensions}

As soon as models aim at some level of realism, they hit a major stumbling block as the number of parameters become large. This means that both their exploration and their calibration are marred with the curse of dimensionality. Instead of getting lost in the wilderness of bounded rationality, ABM run the risk of getting lost in high dimensional spaces. 
Even the heavily stylized Mark-0 model discussed above contains a dozen free parameters \cite{karl_inflation} -- the phase diagram of Mark-0 thus lives in a space that cannot be easily spanned, and the different phases cannot be easily be located. 

However, the pleasant surprise is that in many fields of science (physics, biology, etc.), models are such that only a few ``stiff'' directions in parameter space significantly change the
observable dynamics \cite{gutenkurst,quinn}. Conversely, there are many ``sloppy'' directions in which there are no significant
changes in the model dynamics. The stiff and sloppy directions are defined as the eigenvectors of the
Hessian matrix of a loss function (MSE or KL) around a given point in parameter space, when the log-value of parameters are varied. 

Stiff directions correspond to large eigenvalues. It turns out that these eigenvalues drop very quickly with their rank. This means that most of the model sensitivity to parameters comes from a handful (typically two or three) linear combinations of parameters. This has two major consequences:
\begin{itemize}
    \item Following stiff directions allows an efficient exploration of high dimensional phase diagrams, avoiding to spend time along directions where the outcomes of the model hardly change;
    \item Following sloppy directions until they become zero or infinity allows one to reduce the dimensionality of the model in a non-trivial way \cite{quinn}.
\end{itemize}
In Ref. \cite{karl_sloppy}, we have shown that both agent-based models and DSGE models have a sloppy phenomenology, where their dynamics depend only on a few key stiff parameter directions. In the case of the Mark-0 agent-based model, the stiff parameter
directions point towards close phase transitions. Exploiting these key directions, we developed
an algorithm to systematically explore the phase space of agent-based models, which should apply universally \cite{karl_thesis}. 

\begin{figure}[htb!]
    \centering
    \includegraphics[width=0.7\linewidth]{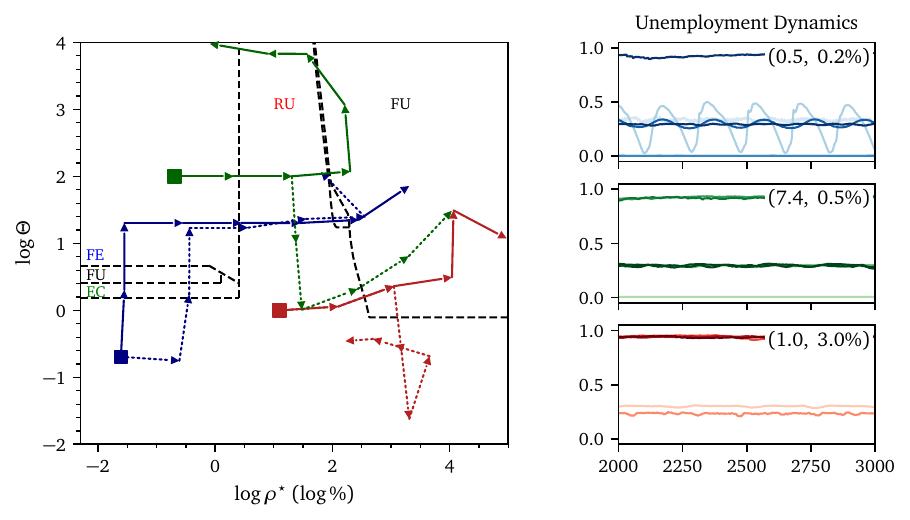}
    \caption{\small{{\it Left:} Phase diagram of the Mark-0 model without CB in the $\Theta,\rho^\star$ plane with two likely algorithm paths for three different starting values. Solid lines show the most likely path following only the top eigenvector ${v}_1$, while dotted lines indicate an alternate path with mixed steps. Dashed black lines indicate the locations of phase transitions. {\it Right:} Dynamics of the unemployment rate for different steps along the ${v}_1$ path for each starting point. The parameters of the model, other than $\Theta$ and $\rho^\star$, are those of Ref. \cite{gualdi}. See \cite{karl_sloppy} for details.} }
    \label{fig:mark0_2d_walks}
\end{figure}

Applying this
algorithm to the Mark-0 model, we have recovered the four phases FU, FE, RU, EC discussed above and identified in \cite{gualdi,monetary}, see Fig. 5, where $\Theta$ is, as above, the bankruptcy threshold and $\rho^\star$ the CB baseline interest rate. This method of exploration also appears to be more computationally efficient as the number of parameter dimensions grows.

Since agent-based models are able to generate a rich phenomenology, such a tool may aid in understanding
the possible scenarios in an agent-based model, as well as their robustness to changes in the
underlying parametrization. This suggests several avenues of further exploration, including extending
our analysis to a wider variety of observable outcomes and more complex agent-based models. Another direction would be to follow sloppy directions instead of stiff directions, with the aim of removing irrelevant (combination of) parameters and constructing minimal models in a systematic way once the different behaviors have been classified, along the lines of \cite{quinn}. Building ``reduced form” models using such a systematic procedure
appears to us as a particularly exciting perspective.

\section{Summary -- Conclusion}

We strongly believe that establishing the phase diagram of ABMs is a crucial first step, together with a qualitative understanding of how collective phenomena come about, before any calibration or more quantitative predictions are attempted. 
Computer-aided {\it gedanken} experiments are by themselves of genuine value: if we are not able to make sense of emergent phenomena in a world in which we set all the rules, how can we expect to be successful in the real world? ABMs indeed often reveal the existence of Black Swans/Dark Corners i.e. discontinuity lines beyond which runaway instabilities appear, whereas most classical economic/finance models are blind to such scenarii.
Testing for the overall robustness of the phase diagram against changes in heuristic rules is a way to ascertain the plausibility of such scenarii. Furthermore, exploring the phase diagrams of ABM in high dimensions should benefit enormously from the identification of ``stiff'' and ``sloppy'' directions in parameter space. 

A qualitative scenario-based approach is often the best one can ever do when dealing with complex systems, except for relatively short term predictions. Such an approach is still useful to help
building faithful intuition and inspiring narratives describing the relevant mechanisms. These are often not even present in DSGE models -- for example, how can one really describe inflation dynamics when markets always clear and when wage bargaining and market power are absent (parameters $g_p$ and $g_w$ mentioned above \cite{karl_inflation})?

Indeed, potentially catastrophic scenarii (like hyper-inflation or default contagion) should be the focus point of policy makers and risk managers, rather than calibrating a bad model to the second decimal digit. As Keynes emphasized long ago, it is better to be roughly right than precisely wrong.

\subsection*{Acknowledgments} I want to thank all my collaborators on these topics, without whom my understanding would be nowhere: M. Benzaquen, T. Dessertaine, S. Gualdi, M. Knicker,  J. Moran, K. Naumann-Woleske, D. Sharma, M. Tarzia, F. Zamponi. I also wish to thank D. Farmer, A. Kirman, A. Mandel, M. Pangallo, A. Roventini \& J. Sethna for many enlightning discussions, and G. Dosi for inviting me to participate to his amazing and exciting 70th birthday festschrift.

\end{document}